\documentclass[12pt]{iopart}
\usepackage{iopams} 
\usepackage{graphicx}

\begin{document}

\title[Faddeev-Merkuriev integral equations for resonances]{Faddeev-Merkuriev integral equations  for atomic three-body resonances}

\author{S Keller, A Marotta and Z Papp}

\address{Department of Physics and Astronomy,
California State University Long Beach, Long Beach, California, USA}

\begin{abstract}
Three-body resonances in atomic systems are calculated as complex-energy solutions of Faddeev-type integral equations. The homogeneous Faddeev-Merkuriev integral equations are solved by approximating the potential terms in a Coulomb-Sturmian basis. The Coulomb-Sturmian matrix elements of the three-body Coulomb Green's operator has been calculated as a contour integral of two-body Coulomb Green's matrices. This approximation casts the integral equation into a matrix equation and the complex energies are located as the complex zeros of the Fredholm determinant. We calculated resonances of the $e-Ps$  system at higher energies and for total angular momentum  $L=1$ with natural and unnatural parity.
\end{abstract}

\submitto{\JPB}

\section{Introduction}

The wave function of a three-particle system is very complicated. It may have several different kinds of asymptotic behavior reflecting the possible asymptotic fragmentations. It is very hard to impose all the asymptotic conditions on a single wave function. The Faddeev approach is a simplification: the wave function is split into components such that each component describes only one kind of asymptotic fragmentation \cite{fm-book}. Then only one kind of asymptotic behavior should be imposed on each component. The components satisfy a set of coupled equations, the Faddeev equations.

If we want to apply this idea to systems with Coulomb potentials, we may run into difficulties. The Coulomb potential is a long range potential, thus the motion in a Coulomb field never becomes a free motion, even at asymptotic distances. Consequently the separation of the wave function along different asymptotic properties does not really work. If we just plug the Coulomb potential into the original Faddeev equations, the equations become singular. The usual asymptotic analysis fails to provide the boundary condition. In integral equation form, the kernel of the equations fails to be compact and we cannot approximate them by finite-rank terms. Merkuriev proposed \cite{fm-book,merkur} a modification of the Faddeev procedure which led to integral equations with compact kernels and differential equations with known boundary conditions. 

Resonances are related to the outgoing-wave solutions of the Schr\"odinger equation at complex energies $E=E_{r}-\mathrm{i}\Gamma/2$. Here $E_{r}$ is the resonance energy and $\Gamma$ is the resonance width, which is related to the lifetime of the decaying state. In an integral equation formalism, the resonances are the solutions of the homogeneous integral equations on the unphysical sheet, close to real energies.

A few years ago, a method for solving Faddeev-type integral equations for scattering problems \cite{scatt3} was adopted to calculate resonances \cite{respra}. The homogeneous version of the Faddeev integral equations was solved at complex energies on the unphysical sheet. The method entails expanding the potentials terms in the integral equations on a Coulomb-Sturmian basis. This transforms the integral equations to a matrix equation. $S$-wave resonances of the $e-Ps$ system have been calculated and  good agreement with the results of methods based on complex rotation of the coordinates were found \cite{ho,shakeshaft}. 

An accumulation of resonance poles around thresholds has been reported in Refs. \cite{prl}, however, we are not investigating these threshold resonances here. They are either too close to the threshold,  too close to each other, or too broad as we move away from the threshold. In any case, their experimental verification does not seem to be likely in the near future.

In this paper, we will report some new developments of this method. In Sec.\ 2, we will outline the Faddeev-Merkuriev approach to the three-body Coulomb problem. In Sec.\ 3, we detail the Coulomb-Sturmian separable expansion approach. We introduce a new contour integral for the three-body Green's operator which makes the analytic continuation to the resonance region easier. In Sec.\ 4 we present resonances of the $e-Ps$  three-body system up to the fifth threshold with total angular momentum  $L=1$.  We provide results for both natural and unnatural parity states.

\section{Faddeev-Merkuriev integral equations}

The  Hamiltonian of an atomic three-body system is given by
\begin{equation}
H=H^0 + v_1^C + v_2^C + v_3^C,
\label{H}
\end{equation}
where $H^0$ is the three-body kinetic energy 
operator and $v_\alpha^C$ denotes the Coulomb
interaction of each subsystem $\alpha=1,2,3$. 
Throughout, we use the usual configuration-space Jacobi coordinates
 $x_\alpha$ and $y_\alpha$, where $x_\alpha$ is the distance
between the pair $(\beta,\gamma)$ and $y_\alpha$ is the
distance between the center of mass
of the pair $(\beta,\gamma)$ and the particle $\alpha$.
Thus, the potential $v_\alpha^C$, the interaction of the
pair $(\beta,\gamma)$, appears as $v_\alpha^C (x_\alpha)$.
In an atomic three-body system, two particles always have the same sign of charge. 
So, without loss of generality, we can assume that they are particles $1$ and $2$, and therefore $v_{3}^{C}$ is a repulsive Coulomb potential.

The  Hamiltonian (\ref{H}) is defined in the three-body 
Hilbert space. Therefore, the two-body potential operators are formally
embedded in the three-body Hilbert space,
\begin{equation}
v^C_\alpha = v^C_\alpha (x_\alpha) {\bf 1}_{y_\alpha},
\label{pot0}
\end{equation}
where ${\bf 1}_{y_\alpha}$ is a unit operator in the 
two-body Hilbert space associated with the $y_\alpha$ coordinate. 

\begin{figure}
\includegraphics[width=0.95\textwidth]{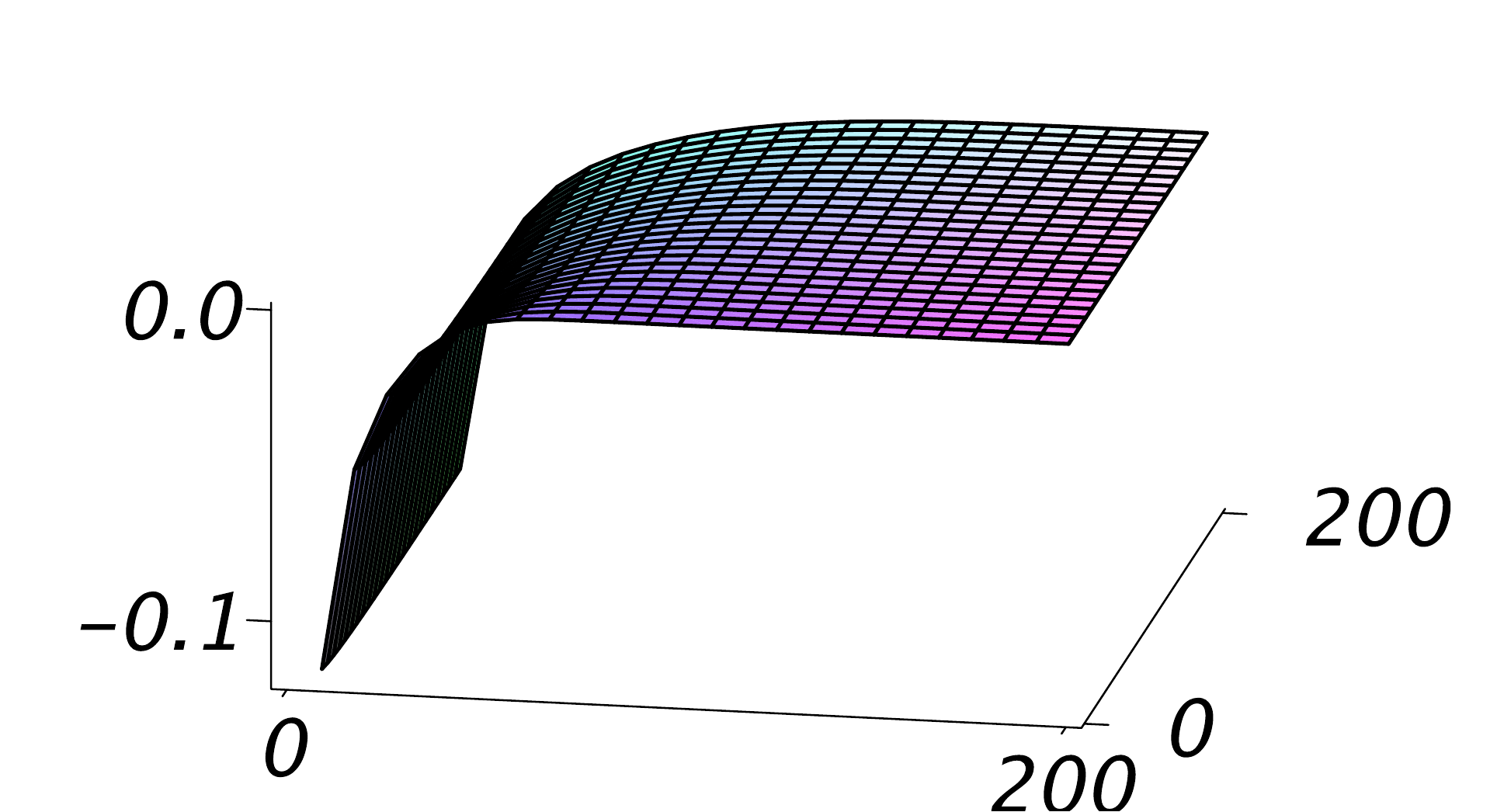}
\caption{Short range potential $v^{(s)}$ for attractive Coulomb
potential. The  parameters are $Z=-1$, $x_{0}=30$, $y_{0}=35$ and $\nu=2.1$ (in atomic units). }
\label{vs}
\end{figure}

\begin{figure}
\includegraphics[width=0.95\textwidth]{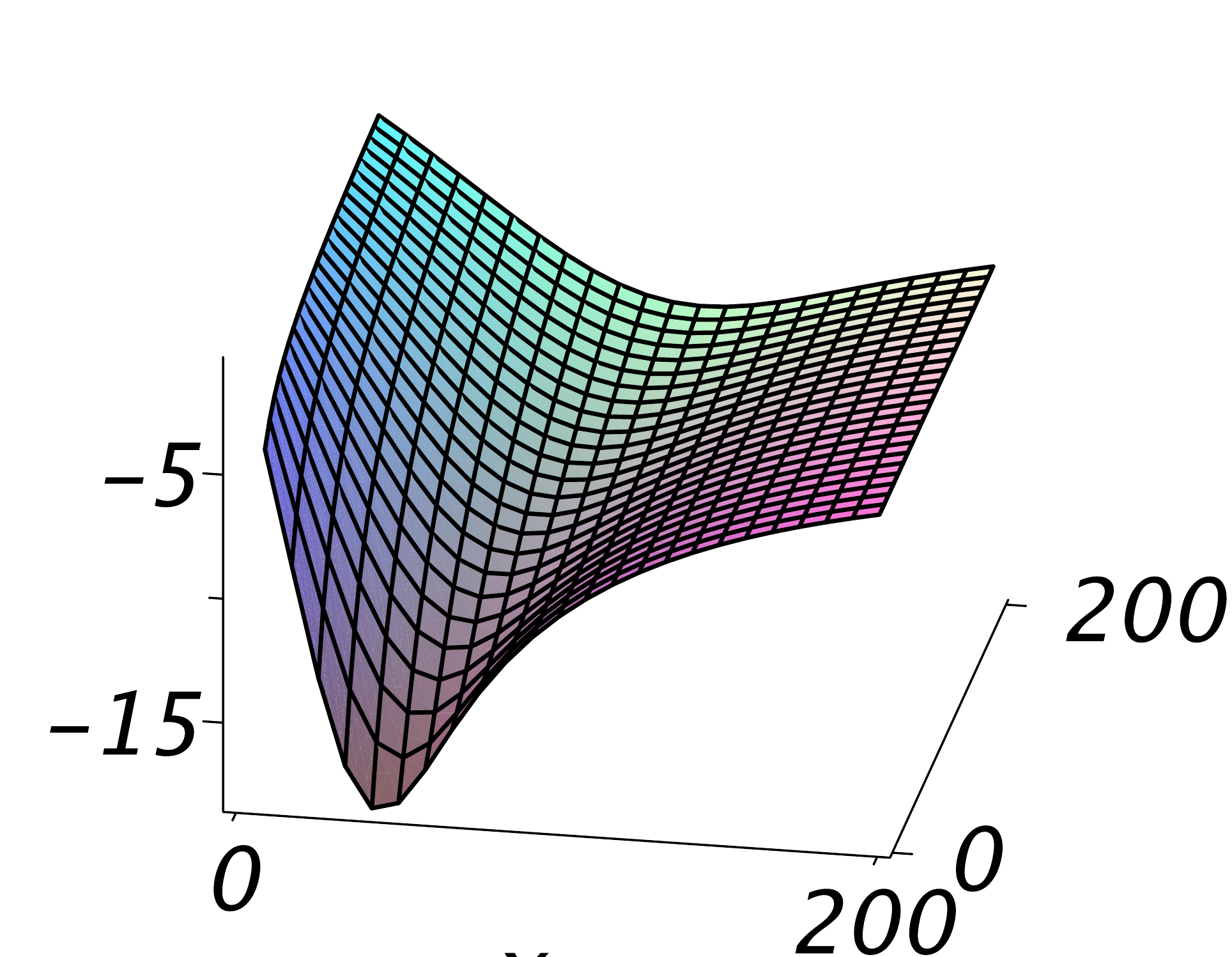}
\caption{Long range potential $v^{(l)}$ for attractive Coulomb
potential. The parameters are as in Fig. 1.}
\label{vl}
\end{figure}

The role of a Coulomb potential in a three-body
system is twofold. The Coulomb potential is a long range potential but it also possesses some features of a short-range potential. It strongly correlates the 
particles and may even support two-body bound states. These two properties are contradictory and require different treatment. 
Merkuriev proposed a separation of the three-body
configuration space into different asymptotic regions \cite{merkur}. 
The two-body asymptotic region $\Omega_\alpha$ is
defined as a part of the three-body configuration space where
the conditions
\begin{equation}
(|x_\alpha|/  x_{0})^{\nu} <   |y_\alpha|/ y_{0},
\label{oma}
\end{equation}
with parameters $x_{0}>0$, $ y_{0} >0$ and $\nu > 2$ are satisfied. 
It was shown that in $\Omega_{\alpha}$ the short-range character of the Coulomb potential prevails, while in the complementary region the long-range character of the Coulomb potential becomes dominant. Thus, it seems to be a good idea to split the Coulomb potential  in 
the three-body configuration space into
short-range and long-range terms 
\begin{equation}
v^C_\alpha =v^{(s)}_\alpha +v^{(l)}_\alpha ,
\label{pot}
\end{equation}
where the superscripts
$s$ and $l$ indicate the short- and long-range
attributes, respectively. 
The splitting is carried out with the help of a splitting function 
$\zeta_\alpha$,
\begin{eqnarray}
v^{(s)}_\alpha (x_\alpha,y_\alpha) & = & v^C_\alpha(x_\alpha) 
\zeta_\alpha (x_\alpha,y_\alpha),
\\
v^{(l)}_\alpha (x_\alpha,y_\alpha) & = & v^C_\alpha(x_\alpha) 
\left[1- \zeta_\alpha (x_\alpha,y_\alpha) \right].
\label{potl}
\end{eqnarray}
The function  $\zeta_\alpha$  
vanishes asymptotically within the three-body sector,
where $x_\alpha\sim y_\alpha \to \infty$, and approaches $1$ in 
the two-body asymptotic region
$\Omega_\alpha$, where $x_\alpha << y_\alpha \to \infty$. 
As a result, in the three-body sector, $v^{(s)}_\alpha$
vanishes and $v^{(l)}_\alpha$ approaches $v^{C}_\alpha$.
In practice, the functional form
\begin{equation}
\zeta_\alpha (x_\alpha,y_\alpha) =  
2/\left\{1+ \exp 
\left[ {(x_\alpha/x_{0})^\nu}/{(1+y_\alpha/y_{0})} 
\right] \right\}
\label{zeta}
\end{equation}
is used. Typical shapes for $v^{(s)}$ and $v^{(l)}$ 
are shown in Figures \ref{vs} and \ref{vl}, respectively. 

In the Hamiltonian (\ref{H}) the Coulomb potential $v_3^C$ is repulsive and does not support bound states. Consequently, there are no two-body channels associated with this 
fragmentation and the entire $v_3^C$ can be considered as long-range potential. Then
the long-range Hamiltonian is defined as
\begin{equation}
H^{(l)} = H^0 + v_1^{(l)}+ v_2^{(l)}+ v_3^{C},
\label{hl}
\end{equation}
and the three-body Hamiltonian takes the form
\begin{equation}
H = H^{(l)} + v_1^{(s)}+ v_2^{(s)}.
\label{hll}
\end{equation}
This Hamiltonian looks like an ordinary three-body Hamiltonian with two
short range interactions. 

To determine the bound and resonant states, we have to solve the 
Schr\"odinger equation
\begin{equation}
H|\Psi\rangle =E|\Psi\rangle
\end{equation}
for real and complex $E$ eigenvalues, respectively. In the Faddeev approach the Faddeev components are defined by
\begin{equation}\label{psidef}
|\psi_{\alpha}\rangle = (E-H^{(l)})^{-1}v_{\alpha}^{(s)} |\Psi \rangle,
\end{equation}
where $\alpha=1,2$.
This involves a splitting of the wave function into two components
\begin{equation}\label{psi12}
|\Psi \rangle =| \psi_1 \rangle + | \psi_2 \rangle.
\end{equation}
Then for the Faddeev components, we have the set of equations, the Faddeev equations,
\begin{eqnarray}
(E-H_{1}^{(l)}) | \psi_1 \rangle &= &  v_1^{(s)} 
| \psi_2  \rangle  \label{fm2c1d}\\
(E-H_{2}^{(l)}) | \psi_2 \rangle &= &  v_2^{(s)} | \psi_1 \rangle,
\label{fm2c2d}
\end{eqnarray}
where 
\begin{equation}\label{halpha}
H_\alpha^{(l)}=H^{(l)}+v_\alpha^{(s)}.
\end{equation} 
By adding these two equations and taking into account Eq.\ (\ref{psi12}) we recover the original 
Schr\"odinger equation. So, the Faddeev procedure is no more and no less than a method of solving the Schr\"odinger equation.
We can cast these differential equations into an integral equation form
\begin{eqnarray}
| \psi_1 \rangle &= & G_1^{(l)} (E) v_1^{(s)} 
| \psi_2  \rangle  \label{fm2c1}\\
| \psi_2 \rangle &= & 
G_2^{(l)}(E) v_2^{(s)} | \psi_1 \rangle,
\label{fm2c2}
\end{eqnarray}
where $G_\alpha^{(l)}(E)=(E-H_{\alpha}^{(l)})^{-1}$.  

Before going further, we should examine the spectral properties of the  Hamiltonian 
\begin{equation}
H_1^{(l)}=H^{(l)}+v_1^{(s)}=H^0+v_1^C+v_2^{(l)}+v_3^C.
\end{equation}
It is obvious that it supports infinitely many two-body channels associated with the 
bound states of the attractive Coulomb potential $v_1^C$. Potential  $v_3^C$ is repulsive, therefore does not support bound states and there are no two-body channels associated with fragmentation $3$.
The three-body potential $v_2^{(l)}$ is attractive.
It is a valley along a parabola-like curve which becomes shallower and shallower, and finally disappears as  $y_2$ goes to infinity (see Fig. \ref{vl}).  Thus, $v_2^{(l)}(x_2,y_2)$ does not support two-body bound states either in the subsystem $x_2$ if $y_2 \to \infty$. 
Consequently, there are no
two-body channels associated with fragmentation $2$.
Therefore, the asymptotic Hamiltonian $H_1^{(l)}$ has two-body channels only in the fragmentation where particle $1$ is at infinity and particles $2$ and $3$ form bound states. If either particle $2$ or $3$ is at infinity, no bound states are allowed in the respective subsystem. The corresponding  $G_1^{(l)}$ Green's operator, acting on the $v_1^{(s)} 
| \psi_2  \rangle$ term in (\ref{fm2c1}), 
will generate only those type of two-body channels in $|\psi_1\rangle$ where particle $1$ is at infinity and particles $2$ and $3$ form bound states.
A similar analysis is valid also for $|\psi_2\rangle$. 
So, the Merkuriev procedure results in a separation of the three-body wave function into
components in such a way that each component has only one type of two-body
channel. This is the main advantage of the original Faddeev equations and,
as the above analysis shows, this property remains valid also for attractive 
Coulomb potentials.

The long-range part of the Coulomb potential, $v^{(l)}_{\alpha}$, does not support two-body channels. It may, however, support bound states, i.e.\ $H^{(l)}$ may have three-body bound states. This can lead to the appearance of spurious solutions of the Faddeev-Merkuriev equations. If $H^{(l)}$ has a bound state, then $(z-H^{(l)})^{-1}$ is singular at this energy. Consequently, in Eq.\ (\ref{psidef}), applying this singular operator on a vanishing $|\Psi\rangle$ may produce a non-vanishing $|\psi_{\alpha}\rangle$. So, we may find a solution where neither $|\psi_{1}\rangle$ nor $|\psi_{2}\rangle$ vanish, but $|\Psi\rangle=|\psi_{1}\rangle + |\psi_{2}\rangle$ vanishes. These states would be non-trivial solutions of the Faddeev-Merkuriev equations, but would be trivial solutions of the original Schr\"odinger equation. These states are spurious, or ghost, solutions. A way to eliminate them is to ensure that in the energy range of physical interest $H^{(l)}$ does not have bound states. We can achieve this by choosing the parameters $x_{0}$ and $y_{0}$ accordingly. For resonances at higher energies we should take a bigger $x_{0}$, thus pushing the unwanted bound states of $H^{(l)}$ out of the spectrum of physical interest. It is also obvious from this analysis that the spurious solutions are sensitive to the choice of $x_{0}$ and $y_{0}$, while the true resonances are not. By varying the parameters $x_{0}$ and $y_{0}$, one can single out the possible spurious solutions.

A very nice advantage of the Faddeev equations is that the identity of particles simplifies the equations. If  particles $1$ and $2$ are identical particles,  the Faddeev components 
$| \psi_1 \rangle$ and $| \psi_2 \rangle$, in their own natural Jacobi
coordinates, must have the same functional forms
\begin{equation}
\langle x_1 y_1 | \psi_1 \rangle = \langle x_2 y_2 | \psi_2 \rangle.
\end{equation}
On the other hand, by interchanging particles $1$ and $2$, we have
\begin{equation}
{\mathcal P_{1 2}} | \psi_1 \rangle = p | \psi_2 \rangle,
\end{equation}
where  $p=\pm 1$ .
Building this information into the formalism, we arrive at
the integral equation
\begin{equation} \label{fmp}
| \psi_{1} \rangle =   G_1^{(l)} v_1^{(s)} p {\mathcal P_{1 2}} 
| \psi_{1} \rangle,
\end{equation}
which by itself determines $| \psi_{1} \rangle$.
We notice that so far no approximation has been made, and even though 
this integral equation has only one component, it
gives a full account of the asymptotic and symmetry properties of the system.

\section{Separable expansion solution of the Faddeev equations}

\subsection{Coulomb-Sturmian basis}

The Coulomb--Sturmian (CS) functions \cite{rotenberg} are 
the solutions of the Sturm--Liouville problem of the Coulomb
Hamiltonian
\begin{equation}
\left( -\frac{\mbox{d}^2 }{ \mbox{d} r^2}
+ \frac{l(l+1)}{ r^2}  - \frac{2b(n+l+1)}{ r}  + b^2 \right)
\langle r\vert n l;b \rangle=0,
\label{sturm-liouville}
\end{equation}
where $b$ is a parameter, $n$ is the radial quantum number and $l$ is the angular momentum.
In configuration space, the CS functions are given by
\begin{equation}
\langle r\vert n l ;b  \rangle 
= \sqrt{ \frac{n!}{(n+2l+1)!} } \ 
\exp(-b r) (2b r)^{l+1} L_n^{(2l+1)}(2b r)\ , 
\label{csf}
\end{equation}
where  $L$ denotes the Laguerre polynomials.
By defining the functions 
$\langle r\vert \widetilde{n l;b} \rangle \equiv \langle r\vert n l;b \rangle/r$, 
the orthogonality and completeness relations take the forms
\begin{equation} \label{orto}
\langle \widetilde{ n'l;b }  \vert nl;b \rangle 
= \langle n'l;b  \vert \widetilde{ n l;b }  \rangle = \delta_{n n'}
\end{equation}
and 
\begin{equation} \label{complet}
{\bf 1} = \lim_{N\to \infty} \sum_{n=0}^N \vert  \widetilde{ n l;b} \rangle 
\langle n l;b \vert 
= \lim_{N\to \infty} \sum_{n=0}^N  \vert  { n l;b} \rangle 
\langle \widetilde{n l;b}  \vert \ .
\end{equation}

Since the three-body Hilbert space is a direct product of two-body
Hilbert spaces, an appropriate basis is the bipolar basis, which
can be defined as the
angular-momentum-coupled direct product of the two-body bases, 
\begin{equation}
| n \nu  l \lambda ; b_x b_y \rangle_\alpha =
 | n  l ; b_x \rangle_\alpha \otimes |
\nu \lambda ;b_y\rangle_\alpha, \ \ \ \ (n,\nu=0,1,2,\ldots),
\label{cs3}
\end{equation}
where $| n  l ;b_x\rangle_\alpha$ and $|\nu \lambda ;b_y\rangle_\alpha$ 
are associated
with the coordinates $x_\alpha$ and $y_\alpha$, respectively.
With this basis, the completeness relation
takes the form (with angular momentum summation implicitly included)
\begin{equation}
{\bf 1} =\lim\limits_{N\to\infty} \sum_{n,\nu=0}^N |
 \widetilde{n \nu l \lambda ;b_x b_y} \rangle_\alpha \;\mbox{}_\alpha\langle
{n \nu l \lambda ;b_x b_y} | =
\lim\limits_{N\to\infty} {\bf 1}^{N}_\alpha,
\end{equation}
where $\langle x y | \widetilde{ n \nu l \lambda;b_x b_y}\rangle = 
\langle x y | { n \nu l \lambda; b_x b_y}\rangle/(x y)$.

\subsection{Separable approximation}

We may introduce an unit operator into the Faddeev equation
\begin{eqnarray}
| \psi_1 \rangle &= & \lim_{N\to\infty } G_1^{(l)} (E) {\bf 1}^{N}_{1 } v_1^{(s)} 
 {\bf 1}^{N}_{2 } | \psi_2  \rangle  \label{fm2c11}\\
| \psi_2 \rangle &= &  \lim_{N\to\infty } G_2^{(l)}(E)  {\bf 1}^{N}_{2 } v_2^{(s)} {\bf 1}^{N}_{1 } | \psi_1 \rangle.
\label{fm2c22}
\end{eqnarray}
This identity becomes an approximation if we keep $N$ finite, which is the equivalent of approximating $v_{\alpha}^{(s)}$  in the three-body
Hilbert space  by a separable form
\begin{eqnarray}
v_{\alpha}^{(s)}  & & =  \lim_{N\to\infty} 
{\bf 1}^{N}_{\alpha} v_\alpha^{(s)}  {\bf 1}^{N}_\beta  \approx 
{\bf 1}^{N}_\alpha v_\alpha^{(s)}  {\bf 1}^{N}_\beta  \approx \nonumber \\ 
& & 
\sum_{n,\nu ,n', \nu'=0}^N
|\widetilde{n\nu l \lambda,b_x b_y}\rangle_\alpha \; \underline{v}_{\alpha \beta}^{(s)}
\;\mbox{}_\beta \langle \widetilde{n' \nu' l' \lambda';b_x b_y}|,  \label{sepfe}
\end{eqnarray}
where $\underline{v}_{\alpha \beta}^{(s)}=\mbox{}_\alpha \langle n\nu l \lambda; b_x b_y |
v_\alpha^{(s)}  |n' \nu' l' \lambda' ;b_x b_y \rangle_\beta$.
These matrix elements can be evaluated numerically
by using the transformation of the Jacobi coordinates \cite{bb}.
The completeness of the CS basis guarantees the convergence of the expansion
with increasing $N$ and angular momentum channels.

Now, by applying the bra $\langle \widetilde{ n'' \nu'' l'' \lambda'';b_x b_y}|$
 from the left, the solution of the homogeneous
Faddeev--Merkuriev equation
turns into the solution of a matrix equation for the component vector
\begin{eqnarray}\label{fm2comp}
\underline{\psi}_1  &= & \underline{G}_1^{(l)} (E) \underline{ v}_{1 2}^{(s)} 
  \underline{\psi}_2    \label{fm2c13}\\
\underline{ \psi}_2  &= &  \underline{G}_2^{(l)}(E)  \underline{ v}_{2 1}^{(s)} \underline{\psi}_1,
\label{fm2c24}
\end{eqnarray}
where
\begin{equation}
\underline{G}_{\alpha}^{(l)}=\mbox{}_\alpha \langle \widetilde{
n\nu l\lambda;b_x b_y} |G_\alpha^{(l)}|\widetilde{n' \nu' l' \lambda';b_x b_y}
\rangle_\alpha.
\end{equation}
These equations can be transformed into the matrix form
\begin{equation}
\left[ 
\left( \matrix{ \underline{1} & \underline{0} \cr \underline{0} & \underline{1} } \right) -
 \left( \matrix{ \underline{G}_{1}^{(l)}(E) & \underline{0} \cr  \underline{0} & \underline{G}_{2}^{(l)}(E) }\right)  
\left( \matrix{ \underline{0} & \underline{v}_{12}  \cr  \underline{v}_{21}  & \underline{0} } \right) \right]
 \left(  \matrix{ \underline{\psi}_{1} \cr \underline{\psi}_{2} } \right)=0,
\end{equation}
which exhibits a homogeneous algebraic equation for the Faddeev components.
This homogeneous algebraic equation is solvable  if and only if 
\begin{equation}
D(E)=\det  \left[ 
 \left( \matrix{ \underline{G}_{1}^{(l)}(E) & \underline{0} \cr  \underline{0} & \underline{G}_{2}^{(l)}(E) }\right)^{-1}  -
\left( \matrix{ \underline{0} & \underline{v}_{12}  \cr  \underline{v}_{21}  & \underline{0} } \right) \right] = 0, \label{homdet}  
\end{equation}
where $D(E)$ is the Fredholm determinant.
The real-energy solutions provide us with the bound states, while the complex-energy ones give the resonant states.

\subsection{Calculation of  $\underline{G}_{\alpha}^{(l)}$}

Unfortunately, the Green's operator  $\underline{G}_\alpha^{(l)}$ 
is not known. It is related to the asymptotic Hamiltonian $H_\alpha^{(l)}$, which is still a complicated three-body Coulomb Hamiltonian. However, $H_{\alpha}^{(l)}$ has  only one type of two-body asymptotic channels where particle $\alpha$ is at infinity. This asymptotic Hamiltonian is denoted by $H_\alpha^{as}$. If a three-body system has only one type of asymptotic channel, then a single Lippmann-Schwinger equation provides a unique solution:
\begin{equation}
G_{\alpha}^{(l)}(z)=G_\alpha^{as}(z) + G_\alpha^{as}(z) V^{as}_\alpha G_\alpha^{(l)}(z),
\label{LSass}
\end{equation}
where $G_1^{as}$ is an asymptotic channel Green's operator, 
$G_{\alpha}^{as}(z)=(z-H_{\alpha}^{as})^{-1}$, and $V^{as}_\alpha=H_{\alpha}^{(l)}-H_{\alpha}^{as}$. Merkuriev constructed $G_\alpha^{as}$ in the different asymptotic regions
of the three-body configuration space
and showed that the kernel of this Lippmann-Schwinger equation 
is completely continuous (compact) \cite{fm-book,merkur}. 
Therefore, $V^{as}_\alpha$ can also be approximated by a separable form 
\begin{eqnarray}
V_\alpha^{as}  & = & \lim_{N\to\infty} 
{\bf 1}^{N}_\alpha  V_\alpha^{as} {\bf 1}^{N}_\alpha 
\approx  {\bf 1}^{N}_\alpha V_\alpha^{as}  {\bf 1}^{N}_\alpha \nonumber \\ 
& \approx  & \sum_{n,\nu ,n', \nu'=0}^N
|\widetilde{n\nu l \lambda;b_x b_y}\rangle_\alpha \; \underline{V}_\alpha^{as}
\;\mbox{}_\alpha \langle \widetilde{n' \nu' l' \lambda';b_x b_y}|,  \label{sepfeas}
\end{eqnarray}
where $\underline{V}_\alpha^{as}=\mbox{}_\alpha \langle n\nu l \lambda;b_x b_y|
V_\alpha^{as} |n' \nu' l' \lambda' ; b_x b_y\rangle_\alpha$.
The solution of Eqs.\ (\ref{LSass})
 can be expressed formally as 
\begin{equation} \label{LSasssol}
(\underline{G}^{(l)}_\alpha )^{-1}= 
(\underline{{G}}^{as}_\alpha )^{-1} - \underline{V}^{as}_\alpha~,
\end{equation}
where 
\begin{equation}
\underline{{G}}^{as}_\alpha =
 \mbox{}_\alpha \langle n \nu l \lambda ;b_x b_y | {G}^{as}_\alpha |
 n' \nu' l' \lambda' ; b_x b_y \rangle_\alpha ,  \label{gasme}
\end{equation}
\begin{equation}
\underline{V}^{as}_\alpha =
 \mbox{}_\alpha \langle n \nu l \lambda; b_x b_y | V^{as}_\alpha | 
 n' \nu' l' \lambda';b_x b_y \rangle_\alpha~. \label{vasme}
\end{equation}

The matrix elements (\ref{gasme}) and (\ref{vasme})
have to be calculated between a finite number
of square-integrable CS states.  In these integrals, the CS functions,
 as a function of $x_{\alpha}$, decay exponentially for large $x_{\alpha}$. 
Hence the domain of integration is confined to $\Omega_\alpha$, where $x_{\alpha}$ is either finite, or $x_{\alpha} << y_{\alpha}$ as $y_{\alpha}\to \infty$. In this region, as Merkuriev showed \cite{merkur}, 
${G}^{as}_\alpha$ takes a simple form; it coincides with the channel Coulomb Green's operator
\begin{equation} \label{gxtilde}
G^{as}_\alpha  = \widetilde{G}_\alpha,
\end{equation} 
where $\widetilde{G}_\alpha(z)=(z- \widetilde{H}_\alpha)^{-1}$, and 
\begin{equation} \label{htilde}
\widetilde{H}_\alpha = H^{0}+v_\alpha^C.
\end{equation}
Therefore, in calculating the matrix elements in Eq.\ (\ref{gasme}), ${G}^{as}_\alpha$
can be replaced by $\widetilde{G}_\alpha$. Similarly, in calculating
(\ref{vasme}), ${V}^{as}_\alpha$
can be replaced by 
\begin{equation}
U_\alpha=v_\beta^{(l)}+v_3^C.
\end{equation}
Consequently, Eq.\ (\ref{LSasssol}) becomes
\begin{equation}
(\underline{G}^{(l)}_\alpha)^{-1}= 
(\underline{\widetilde{G}}_\alpha)^{-1} - \underline{U}_\alpha,
\label{gleq}
\end{equation}
where 
\begin{equation}
\underline{\widetilde{G}}_{\alpha} =
 \mbox{}_\alpha \langle \widetilde{n \nu l \lambda;b_x b_y} | 
 \widetilde{G}_\alpha | \widetilde{ n' \nu' l' \lambda';b_x b_y} \rangle_\alpha
 \label{gtilde}
\end{equation}
and 
\begin{equation}
\underline{U}_{\alpha} =
 \mbox{}_\alpha \langle n\nu l \lambda;b_x b_y | U_\alpha | 
 n' \nu' l' \lambda';b_x b_y \rangle_\alpha.
\end{equation}
The  $\underline{U}_{\alpha}$ matrix elements can again be evaluated numerically.

\subsection{Matrix elements of $\widetilde{G}_{\alpha}$}

The most crucial point in this procedure is the calculation of the matrix elements
$\underline{\widetilde{G}}_\alpha$.
In our Jacobi coordinates, the three-particle free Hamiltonian
can be written  as a sum of two-particle free Hamiltonians 
\begin{equation}
H^0=h_{x_\alpha}^0+h_{y_\alpha}^0.
\end{equation}
Thus the Hamiltonian $\widetilde{H}_\alpha$ of Eq.\ (\ref{htilde}) 
appears as a sum of two two-body Hamiltonians acting on different coordinates 
\begin{equation}
\widetilde{H}_\alpha =h_{x_\alpha}+h_{y_\alpha},
\end{equation}
where $h_{x_\alpha}=h_{x_\alpha}^0+v_\alpha^C(x _\alpha)$ and $h_{y_\alpha}=h_{y_\alpha}^0$, 
which, of course, commute. As a result,  $\widetilde{G}_\alpha$
is a resolvent of the sum of two commuting Hamiltonians $h_{x_{\alpha}}$ and $h_{y_{\alpha}}$.

According to the Dunford-Taylor functional calculus, a function of  a self-adjoint operator $h$ can be defined by 
\begin{equation}
f(h)=\frac{1}{2\pi {i}} \oint_{C} dz'\, f(z') (z'-h)^{-1},
\end{equation}
where $C$ encircles the spectrum of $h$ in counterclockwise direction and $f$ is analytic on the area
encircled by $C$.
In that way, 
$\widetilde{G}_\alpha$, as a function of the self-adjoint operator $h_{x_{\alpha}}$,  can be written as
\begin{eqnarray}
\widetilde{G}_\alpha (z)&=& (z- h_{y_{\alpha}}-h_{x_{\alpha}})^{-1}  \nonumber \\
&=& \frac 1{2\pi {i}}\oint_C  dz' \,(z-h_{y_{\alpha}}-z')^{-1} \;(z'-h_{x_{\alpha}})^{-1}   \nonumber \\
&=& \frac 1{2\pi {i}}\oint_C  dz' \,g_{y_\alpha}(z-z')\;g_{x_\alpha}(z'),
\label{contourint}
\end{eqnarray}
where
$g_{x_\alpha}(z)=(z-h_{x_\alpha})^{-1}$  and
$g_{y_\alpha}(z)=(z-h_{y_\alpha})^{-1}$.
The contour $C$ should be taken in a counterclockwise direction
around the singularities of $g_{x_\alpha}$
such that $g_{y_\alpha}$ is analytic on the domain encircled
by $C$. Accordingly, to calculate the matrix elements $\widetilde{\underline{G}}_\alpha$, we need to calculate a contour integral of the two-body Green's matrices $\underline{g}_{y_\alpha}$ and $\underline{g}_{x_\alpha}$.

In our case, $g_{x_{\alpha}}$ is a Coulomb Green's operator with a branch-cut on the  $[0,\infty)$ interval and accumulation of infinitely many bound states at zero energy, while $g_{y_{\alpha}}$ is a free Green's operator with branch-cut on the $[0,\infty)$ interval.
In time-independent scattering theory, $\widetilde{G}_\alpha (E)$
should be understood as  
$\widetilde{G}_\alpha (E)=\lim_{\varepsilon\to 0} \widetilde{G}_\alpha (E +{\mathrm{i}}\varepsilon)$, with $\varepsilon > 0$. To calculate resonances, we need to continue analytically to $\varepsilon < 0$. In this paper, we limit our study to energies 
below the three-body breakup threshold, so $\Re(E) < 0$.

\begin{figure}[htbp]
\includegraphics[width=0.95\textwidth]{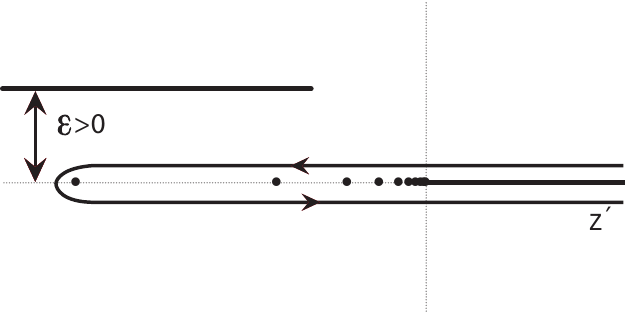}
\caption{The analytic structure of $g_{y_\alpha}(E+{\mathrm{i}}\varepsilon-z')
g_{x_\alpha}(z')$ as a function of $z'$, $\varepsilon>0$. The operator
$g_{x_\alpha}(z')$ has a branch-cut on the $[0,\infty)$ interval and accumulation of infinitely many
bound states at zero energy, while
$g_{y_\alpha}(E+{\mathrm{i}}\varepsilon-z')$ has a branch-cut on the 
$(-\infty,E+{\mathrm{i}}\varepsilon]$ interval.
The contour $C$ encircles the spectrum of
$g_{x_\alpha}$ and avoids the singularities of $g_{y_{\alpha}}$. }
\label{fig1}
\end{figure}

\begin{figure}[htbp]
\includegraphics[width=0.95\textwidth]{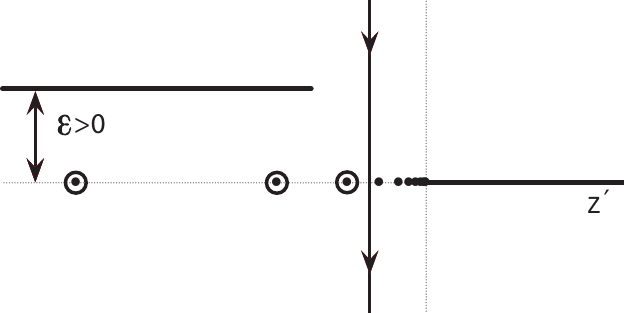}
\caption{The contour of Fig.\ \ref{fig1} is
deformed analytically such that it shrinks to the low-lying bound-state poles of $g_{x_\alpha}$ and the other part is taken along an imaginary direction. }
\label{fig2}
\end{figure}

\begin{figure}[htbp]
\includegraphics[width=0.95\textwidth]{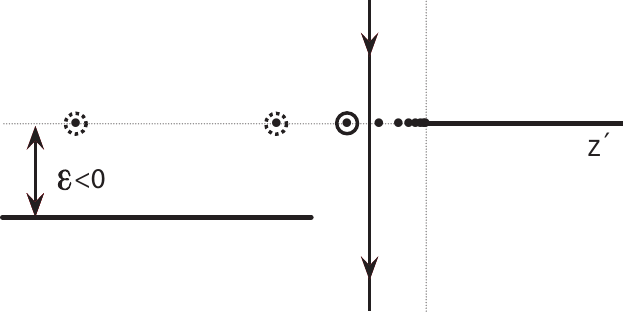}
\caption{Even in the $\varepsilon < 0$ case, which is needed to calculate resonances, the singularities remain separated. Some low-lying poles of $g_{x_{\alpha}}$ submerge onto the second Riemann sheet of $g_{y_{\alpha}}$, and they are denoted by dotted contour.}
\label{fig4}
\end{figure}

To examine the analytic structure of the integrand in Eq.\ (\ref{contourint}) 
let us take $\varepsilon>0$. By doing so,
the singularities of $g_{x_\alpha}$ and $g_{y_\alpha}$ become well separated.
Now the spectrum of $g_{x_\alpha}$ can easily be encircled so  that
the singularities of $g_{y_\alpha}$ lie outside the encircled domain 
(Fig.~\ref{fig1}). However, this would not be the case for $\varepsilon \le 0$.
Therefore the contour $C$ is deformed analytically in
such a way that it shrinks to a few lowest bound states and the contour opens up and continues along an imaginary line (Fig.~\ref{fig2}).  Now,  even in the $\varepsilon < 0$ case (Fig.~\ref{fig4}),
the contour  avoids the singularities of $g_{y_\alpha}$.
Thus, the mathematical conditions for
the contour integral representation of $\widetilde{G}_\alpha$ in
Eq.~(\ref{contourint}) are met also for resonant-state energies. 

\subsection{The Coulomb-Sturmian matrix elements of the Coulomb Green's operator}

In our system, $g_{y_{\alpha}}$ is a free Green's operator and $g_{x_{\alpha}}$ is a Coulomb Green's operator. Their CS matrix elements can be calculated analytically \cite{dhp}.
The two-body Coulomb Green's operator is the resolvent of the Coulomb Hamiltonian
\begin{equation}
g_{l}^{C}=(z-h_{l}^{C})^{-1},$$
\end{equation}where
\begin{equation}
 h^{\rm C}_l=-\frac{\hbar^2}{2m}\left(\frac{\mbox{d}^2 }{\mbox{d} r^2}
- \frac{l(l+1)}{ r^2}\right) + \frac{Z}{ r}\ ,
\label{coulham}
\end{equation}
$m$ is the reduced mass, and $Z$ is the strength of the Coulomb potential.
In the CS basis the operator $J=z-h_{l}^{C}$ has an infinite symmetric tridiagonal (Jacobi) matrix structure, i.e.\ all elements are zero, except for the diagonals and off-diagonals,
\begin{equation}
{J}^{\rm C}_{ii}=2(i+l+1) (k^2-b^2 )
\frac{\hbar^2}{4mb}- Z \ ,
\label{jii}
\end{equation}
\begin{equation}
{J}^{\rm C}_{i i-1}=-[i(i+2l+1)]^{1/2} (k^2+b^2 )
\frac{\hbar^2}{4mb} 
\label{jiip1}
\end{equation}
and
\begin{equation}
{J}^{\rm C}_{i i+1}=-[(i+1)(i+2l+2)]^{1/2} (k^2+b^2 )
\frac{\hbar^2}{4mb} \ , 
\label{jiip2}
\end{equation}
where $k=(2m z/\hbar^2)^{1/2}$.
Then, as it has been shown in Ref.\  \cite{dhp}, the $N\times N$ matrix elements of  ${g}_{l}^{{\rm C}}$ are given by
\begin{equation}
\label{invnu}
\underline{g}_{l}^{{\rm C} (N)} =[\underline{J}^{\rm C} -
\delta _{jN}\, \, \delta _{iN}\, \, ({J}^{\rm C}_{N N+1})^{2} \, 
\, C_{N+1} ]^{-1}\ ,
\end{equation}
where $\underline{J}^{\rm C}$ is the $N\times N$ upper left corner of the Jacobi matrix and
\begin{eqnarray}\label{eq:coulomb-cf-final}
C_{N+1} &=&
- \frac{ \displaystyle  4m/\hbar^{2} b } 
{ \displaystyle    \left({\displaystyle  b  -   \mathrm{i}  k  }\right)^{2} 
\left( \displaystyle  N+l+2 +   \mathrm{i}  \gamma \right)} \nonumber \\
&& \times \frac{{_2}F{_1}\left( \displaystyle - l +   \mathrm{i}  \gamma  , N+2; N+l+3 +   \mathrm{i}  \gamma   ;
\left(\frac{{b}  + \mathrm{i} k  }   {{b} -   \mathrm{i}  k  }\right)^2\right)}
{{_2}F{_1}\left( \displaystyle -l+  \mathrm{i}  \gamma , N+1; N+l+2 +  \mathrm{i}  \gamma ;
\left(\frac{ { b} + \mathrm{i} k }{ { b} -   \mathrm{i} k }\right)^2\right)}~,
\end{eqnarray}
with $_{2}F_{1}$ being the hypergeometric function and $\gamma=Z/(m \hbar^{2}k)$.
This ratio of two   $_{2}F_{1}$ functions, where the second index in the numerator and in the denominator differ by one, can be represented 
by a continued fraction \cite{lorentzen}, which is easily computable and  convergent on the whole complex $k$  plane.

\subsection{Numerical realization of the method}

In this approach for solving the Faddeev-Merkuriev integral equation, the only approximation is the replacement of the potentials $v_{\alpha}^{(s)}$ and $U_{\alpha}$ by their respective separable forms. We found that good results are achieved when we use $N$ up to $25-30$ in the separable expansion for each angular momentum channel. To calculate the matrix elements between CS functions, which are, in fact, exponential functions multiplied by polynomials, we use Gaussian integration;  about $120-150$ points provide the sufficient accuracy.

The calculation of $\underline{\widetilde{G}}_{\alpha}$ is very accurate. It should be noted first that this representation of $\underline{g}_{l}^{\rm C}$ is exact, and its numerical realization, including the evaluation of the ratio of two $_{2}F_{1}$ functions by a continued fraction is precise to machine accuracy. The contour integral around the poles of $g_{x_{\alpha}}$ is a projection onto the corresponding bound state 
 \begin{equation}
|\phi_{i}\rangle \langle \phi_{i}| =\frac{1}{2\pi {i}} \oint_{C_{i}} dz\,  g_{x_{1}}(z)~,
\label{bcontour}
\end{equation}
where $\phi_{i}$ is the eigenstate belonging to the eigenvalue $E_{i}$, and $C_{i}$ is a contour around $E_{i}$. In fact, the states $|\phi_{i}\rangle$ are hydrogenic bound states.  We calculated the overlap $\langle nl;b|\phi_{i}\rangle$  using (\ref{bcontour}) and compared it with the exact result in Maple. We found a perfect agreement.  We also found that the main contribution to $\underline{\widetilde{G}}_{\alpha}$ is due to the bound state poles. The contour integral along the imaginary line behaves asymptotically like $1/(1+z')^{2}$. We adopted the Gauss-rational integration method, and found that about $50-60$ integration points provide a sufficient accuracy.  This is a significant improvement over previous methods which employed as many as $250$ integration points in Refs.\ \cite{respra,prl} to achieve a comparable level of accuracy.
 
In order to find those complex zeros of $D(E)$, which are close to the real energy line,
we have developed the following procedure. We consider an interval along the real energy line between two thresholds. Since $D(E)$ is an analytic function of the energy we can approximate it with Chebyshev polynomials. We use about $n=12-15$ Chebyshev polynomials. The length of the interval should be small enough that $D(E)$ does not change too much and thus the Chebyshev approximation is reliable. The zeros of the Chebyshev approximated function is determined by using the eigenvalue method of Ref.\ \cite{boyd}.
Then the rank of the Chebyshev  approximation is lowered by one, and the zeros are located  again. If a zero is a true zero of $D(E)$, the zeros of the rank $n$ and rank $n-1$ Chebyshev polynomials are close. A similar concept was adopted in Ref.\ \cite{elander} using Pad\`e approximation instead of Chebyshev. We then look for the zeros of $D(E)$ in the neighborhood of the zeros of the Chebyshev approximation. We pick three complex points, $z_{1}$, $z_{2}$ and $z_{3}$. The location $z_{0}$ of the complex root is estimated by \cite{lovas}
\begin{equation}
z_{0}=\frac{z_{1}(z_{2}-z_{3})/D(z_{1})  + z_{2}(z_{3}-z_{1})/D(z_{2}) +  z_{3}(z_{1}-z_{2})/D(z_{3})}
{ (z_{2}-z_{3})/D(z_{1})+ (z_{3}-z_{1})/D(z_{2})+  (z_{1}-z_{2})/D(z_{3})}.
\end{equation}
Then we make a replacement $z_{0}\to z_{1}$, $z_{1}\to z_{2}$ and $z_{2}\to z_{3}$, and repeat until $|z_{0}-z_{1}|< \epsilon$ with some small $\epsilon$. If the initial estimation for the zero is good, this procedure converges very fast. After some experience, we found this method quite fast and reliable.

\section{Results}

We calculated the resonances of the electron-positronium,  $e^{}-Ps$ or $e^{-}-e^{-}-e^{+}$,  three-body system.  Here the two electrons are identical particles, allowing us to use the one-component version of the homogeneous Faddeev-Merkuriev equations (\ref{fmp}). We use atomic units throughout. 
For the parameters of the cut-off function (\ref{zeta}) we adopted $x_0= 30$, $y_0= 35$ and
$\nu=2.1$. This choice of parameters guarantees that in the energy region up to the fifth threshold, there are no spurious solutions. 

The parity of the states is given by $P=(-)^{l+\lambda}$. If $P=(-)^{L}$, the state has natural parity, if $P=(-)^{L+1}$, the state has unnatural parity. The wave function should be antisymmetric with respect to the exchange of the two electrons. If the spin of the two electrons couple to $S=0$ to form a singlet state, then the wave function is antisymmetric with respect to the exchange of electron-spin coordinates, and the spacial part should be symmetric. Similarly, if the two electrons couple to $S=1$ forming a triplet state, the wave function is symmetric with respect to exchange of the spin coordinates, and the spatial part of the wave function is antisymmetric. Consequently, in Eq.\ (\ref{fmp}), if $S=0$ then $p=1$ and if $S=1$ then $p=-1$.

We present results for total angular momentum $L=1$. The angular momentum quantum numbers $l$ and $\lambda$ are selected such that $\vec{l}+\vec{\lambda}=\vec{L}$ . Table \ref{tabone} shows the angular momentum channels used in these calculations. 

 \begin{table}
\caption{\label{tabone}
Angular momentum channels $l-\lambda$ used for  $L=1$ states with natural and unnatural parity. The superscript stands for the parity.} 
\begin{indented}
\lineup
\item[]\begin{tabular}{@{}*{2}{l}}
\br                              
 $L=1^{-}$ & $L=1^{+}$\cr 
\mr
 0-1  & 1-1  \cr
 1-0  & 2-2  \cr
 1-2  & 3-3  \cr
 2-1  & 4-4  \cr
 2-3  & 5-5  \cr
 3-2  & 6-6  \cr
 3-4  & 7-7  \cr
 4-3  & 8-8  \cr
\br
\end{tabular}
\end{indented}
\end{table}

In this method, we represent operators on the CS basis, which has one parameter, the parameter $b$. To be economic, we need to find an optimal $b$, and then we need to 
increase the basis size $N$ to observe convergence. We found that the results are insensitive to varying $b$ over a rather broad interval around $b=0.25$. We used $b=0.25$ throughout. Table \ref{conv} shows a typical convergence of a resonant-state energy with increasing $N$. From results like this, we can safely infer about three significant digits for the real part of energy and one or two significant digits for the imaginary part of the energy. Tables \ref{L1n} and \ref{L1un} show the results of our calculations.  

 \begin{table}
\caption{\label{conv}
The convergence of a resonant-state energy with increasing $N$. The results are given in atomic units. $E_{r}$ is the real part, $E_{i}$ is the imaginary part of the energy.} 
\begin{indented}
\lineup
\item[]\begin{tabular}{@{}*{3}{l}}
\br                              
 $N$ & $E_{r}$ & $E_{i}$ \cr 
\mr
   25  &   -0.028958565292  &   -0.000000304662  \cr
   26   &  -0.028959758422   &  -0.000000303622  \cr
   27   &  -0.028960583605   &  -0.000000302855  \cr
   28   &  -0.028961155544   &  -0.000000302318  \cr
   29   &  -0.028961553239   &  -0.000000301934  \cr
   30   &  -0.028961831103   &  -0.000000301642  \cr
 \br
\end{tabular}
\end{indented}
\end{table}

 \begin{table}
\caption{\label{L1n}
$L=1$ natural parity singlet (S=0) and triplet (S=1) resonances in the $e-Ps$ system. 
The thresholds are indicated by empty lines. The resonance energies $E=E_{r}-\mathrm{i}\Gamma/2 $ are given in atomic units. }
\begin{indented}
\lineup
\item[]\begin{tabular}{@{}*{4}{l}}
\br                              
 $S=0$ & & $S=1$ & \cr

\mr
 $ -0.2340    $&$		-0.0050		\mathrm{i}$	&  $ -0.248    	$&$	-0.003			\mathrm{i}$ \cr
 $ -0.06257  $&$ 	-0.0000000007		\mathrm{i}$	&  $ -0.24     	$&$	-0.007			\mathrm{i}$ \cr
&											&  $ -0.23456	$&$   -0.00263			\mathrm{i}$ \cr
 & \cr
  $-0.0619     	$& $		-0.0005		\mathrm{i}$	&  $ -0.0619   $&$	-0.0005			\mathrm{i}$ \cr
  $-0.059948   $& $  	-0.0001461		\mathrm{i}$	&  $ -0.0611   $&$ 	-0.0009			\mathrm{i}$ \cr
  $-0.0306834 $& $      -0.00006229		\mathrm{i}$	&  $ -0.02926 $&$   -0.0000261		\mathrm{i}$ \cr
  $-0.02896     $& $	-0.00000030		\mathrm{i}$	&  $ -0.0281   $&$ 	-0.000008			\mathrm{i}$ \cr
  &											&  $  -0.02794$&$    -0.00000006		\mathrm{i}$ \cr
 & \cr
  $-0.018479  $& $ 	    -0.00001267		\mathrm{i}$	&  $ -0.0276  $& $ 	-0.0001			\mathrm{i}$ \cr
  $-0.01635    $& $ 	-0.000006			\mathrm{i}$	&  $ -0.01992$& $   -0.000061			\mathrm{i}$ \cr
  $-0.0160      $& $		-0.0000004	\mathrm{i}$	&  $ -0.0173  $& $	-0.000056			\mathrm{i}$ \cr
  $-0.01569    $& $ 	-0.0000001		\mathrm{i}$	&  $ -0.0164  $& $ 	-0.0000007		\mathrm{i}$ \cr
 & \cr
  $-0.01224  $& $	-0.0000063		\mathrm{i}$	&  $ -0.0156   $& $	-0.0000001		\mathrm{i}$ \cr
  $-0.0107    $& $		-0.00001		\mathrm{i}$	&  $ -0.0107   $& $	-0.000001			\mathrm{i}$ \cr
 \br
\end{tabular}
\end{indented}
\end{table}

 \begin{table}
\caption{\label{L1un}
$L=1$ unnatural parity singlet and triplet resonances in the $e-Ps$ system. 
The thresholds are indicated by empty lines. The resonance energies $E=E_{r}-\mathrm{i}\Gamma/2 $ are given in atomic units. }
\begin{indented}
\lineup
\item[]\begin{tabular}{@{}*{4}{l}}
\br                              
 $S=0$ && $S=1$& \cr 
\mr
 $ -0.0610   $& $ 	-0.00158			\mathrm{i}$	&  $ -0.061901    $& $ 	-0.0000993\mathrm{i}$ \cr
 $ -0.061     $& $-0.003				\mathrm{i}$	&  $ -0.060     	 $& $	-0.002			\mathrm{i}$ \cr
 $ -0.02816 $& $   -0.000000002\mathrm{i}$	&  $ -0.031560    $& $	-0.0000866\mathrm{i}$ \cr
&&&\cr
 $ -0.0277  $& $ 	-0.00003			\mathrm{i}$	&  $ -0.0277     	$& $		-0.00006\mathrm{i}$ \cr
 $ -0.027    $& $	-0.0005			\mathrm{i}$	&  $ -0.027     	$& $		-0.0005\mathrm{i}$ \cr
 $ -0.0165  $& $  	-0.00000005	\mathrm{i}$	&  $ -0.018857   $& $		-0.00000690\mathrm{i}$ \cr
 $-0.0157   $& $ 	-0.00000003	\mathrm{i}$	&  $ -0.016     	$& $		-0.000002\mathrm{i}$ \cr
&&&\cr
 $-0.0108    $& $	-0.0000002		\mathrm{i}$	&  $ -0.0125     	$& $		-0.00001\mathrm{i}$ \cr
 $-0.0100    $& $	-0.0000003		\mathrm{i}$	&  $ -0.0108     	$& $		-0.00007\mathrm{i}$ \cr\br
\end{tabular}
\end{indented}
\end{table}

\section{Summary}
In this paper, we outlined a solution method for the homogeneous Faddeev-Merkuriev integral equations to calculate resonances in atomic three-body systems. We approximated the potential terms in the three-body Hilbert space by a separable form. This approximation casts the integral equations into a matrix equation and the resonances are sought as complex-energy roots of the Fredholm determinant. The matrix elements of the three-body channel Coulomb Green's operator were evaluated as a complex contour integral of the two-body Coulomb Green's matrices. The use of the Coulomb-Sturmian basis allows analytic evaluation of these matrix elements. We found that the contour introduced here is more advantageous  than those used in our previous publications \cite{respra,prl}. The method is quite efficient. To achieve  good accuracy we do not need too many terms in the expansion, only  $N=30$ in each angular momentum channel, and consequently the size of the matrix is relatively small. We performed all of our calculations with Mac PC's. We calculated resonances of the $e-Ps$  atomic three-body system for total angular momentum  $L=1$ with natural and unnatural parity.  We do not believe that there is an ultimate method to calculate resonances. However, our results allow us to believe that  this solution of the homogeneous Faddeev-Merkuriev equations is an accurate and reliable method for calculating resonances in atomic three-body systems.

\section{Acknowledgments}
The authors are thankful to S.\ L.\ Yakovlev for useful discussions. This work has been supported by the Research Corporation.

\end{document}